\documentclass[twocolumn,epjc3]{svjour3}  

\usepackage{amssymb}     
\newcommand{\ctsper}      {cts/(keV$\cdot$kg$\cdot$yr)}

\newcommand{\kgy}         {{kg$\cdot$yr}}

    \newcommand{\mus}         {{$\mu$s}}
\newcommand{\onbb}        {{$0\nu\beta\beta$}}

\newcommand{\etal}        {\textit{et al.}}

\newcommand{\gerda}       {\textsc{Gerda}}

\newcommand{\lngs}        {{\mbox{\textsc{Lngs}}}}

\newcommand{\phaseone}    {Phase~I}

\newcommand{\majorana}    {\textsc{Majorana}}

\newcommand{\geant}       {\textsc{Geant4}}

\newcommand{\nuc}[2]      {{$^{#2}$\rm #1}}

\newcommand{\SI}[2]       {{\rm{#1}\,\rm #2}}

\newcommand{\LEG}       {LE\-GEND}
\newcommand{\Ltwo}      {{\LEG-200}}

\newcommand{\Lk}        {{\LEG-1000}}

\newcommand{\germg}     {{$^{77}$Ge}}

\newcommand{\pt}        {{Phase~II}}

\smartqed  
\RequirePackage{graphicx}
\usepackage[switch]{lineno}
\usepackage{url}

\usepackage{verbatim}

\nolinenumbers

\journalname{Eur. Phys. J. C}
\begin{document}

    \title{Search for the in-situ production of $^{77}$Ge in the GERDA neutrinoless double-beta decay experiment}


\author{
The \mbox{\protect{\sc{Gerda}}} collaboration\thanksref{corrauthor}
\and  \\[4mm]
%
M.~Agostini\thanksref{UCL} \and
A.~Alexander\thanksref{UCL} \and
G.~Araujo\thanksref{UZH} \and
A.M.~Bakalyarov\thanksref{KU} \and
M.~Balata\thanksref{ALNGS} \and
I.~Barabanov\thanksref{INRM,deceased} \and
L.~Baudis\thanksref{UZH} \and
C.~Bauer\thanksref{HD} \and
S.~Belogurov\thanksref{ITEP,INRM,alsoMEPHI} \and
A.~Bettini\thanksref{PDUNI,PDINFN}
% ...
}
\authorrunning{the \textsc{Gerda} collaboration}
\thankstext{corrauthor}{
  \emph{correspondence:}  gerda-eb@mpi-hd.mpg.de}
\thankstext{deceased}{\emph{deceased}} 
\thankstext{alsoMEPHI}{\emph{also at:} NRNU MEPhI, Moscow, Russia}
\thankstext{nowDuke}{\emph{present address:} Duke University, Durham, NC USA}
\thankstext{alsoLev}{\emph{also at:} Moscow Inst. of Physics and Technology,
  Russia}
\thankstext{nowZrt}{\emph{present address:} Semilab Zrt, Budapest, Hungary}
\thankstext{alsoDubna}{\emph{also at:} Dubna State University, Dubna, Russia}
\thankstext{nowBerkeley}{\emph{permanent address:} Nuclear Science Division, Berkeley, USA}
\institute{ 
INFN Laboratori Nazionali del Gran Sasso, Assergi, Italy\label{ALNGS}
% ...
\and
Max-Planck-Institut f{\"u}r Kernphysik, Heidelberg, Germany\label{HD} \and
Department of Physics and Astronomy, University College London, London, UK\label{UCL}
% ...
\and
Institute for Nuclear Research of the Russian Academy of Sciences, Moscow, Russia\label{INRM} \and
Institute for Theoretical and Experimental Physics, NRC ``Kurchatov Institute'', Moscow, Russia\label{ITEP} \and
National Research Centre ``Kurchatov Institute'', Moscow, Russia\label{KU}
% ...
\and
Dipartimento di Fisica e Astronomia, Universit{\`a} degli Studi di 
Padova, Padua, Italy\label{PDUNI} \and
INFN  Padova, Padua, Italy\label{PDINFN}
% ...
\and
Physik-Institut, Universit{\"a}t Z{\"u}rich, Z{u}rich, Switzerland\label{UZH}
} 

\date{Received: date / Accepted: date}

\maketitle

\begin{abstract}
The beta decay of {$^{77}$Ge and $^{77\mathrm{m}}$Ge}, both produced by neutron capture on $^{76}$Ge, is a potential background for Germanium based 
neutrinoless double-beta decay search experiments such as \gerda\  or the \LEG\ experiment. 
In this work we present a search for $^{77}$Ge decays in the full \gerda\ Phase~II data set. 
A delayed coincidence method was employed to identify the decay of $^{77}$Ge via the isomeric state of $^{77}$As (9/2$^+$, $\SI{475}{keV}$, 
${T_{1/2} = \SI{114}{\mu s}}$, {$^{77\mathrm{m}}$As}).
New digital signal processing methods were employed to select and analyze pile-up signals.
No signal was observed, and an upper limit on the production rate of $^{77}$Ge was set at $<0.216$\,{nuc/(kg$\cdot$yr)} (90\% CL). 
This corresponds to a total production rate of $^{77}$Ge and $^{77\mathrm{m}}$Ge of $<\SI{0.38}{nuc/(kg\cdot yr)}$ (90\% CL), assuming equal 
production rates. 
A previous Monte Carlo study predicted a value for in-situ \nuc{Ge}{77} and $^{77\mathrm{m}}$Ge production of (0.21$\pm$ 0.07)\,nuc/(kg.yr), a 
prediction that is now further corroborated by our experimental limit.
Moreover, tagging the isomeric state of $^{77\mathrm{m}}$As can be utilised to further suppress the \nuc{Ge}{77} background. 
Considering the similar experimental configurations of \Lk\ and \gerda, the cosmogenic background in \Lk\ at LNGS is estimated to remain at a 
sub-dominant level.
\end{abstract}


\section{Introduction}
\label{intro}

In-situ production of radioactive isotopes by atmospheric muons can represent a non-negligible background for experiments searching for rare events 
even when located deep underground. One such experiment is the \gerda\ ({\sc Ger}\-manium {\sc D}etector {\sc A}rray) 
experiment \cite{gerda:phase2:2018} that searched for the neutrinoless double-beta ($0\nu\beta\beta$) decay in $^{76}$Ge located underground 
below a rock overburden of about 3.5 km.w.e. at the \lngs\ (Laboratori Nazionali del Gran Sasso) of INFN.

In \pt\ of the \gerda\ experiment, 40 (after upgrade 41) high-purity germanium (HPGe) detectors made from material isotopically enriched in 
$^{76}$Ge were operated as bare crystals in a cryostat filled with 64~m$^3$ liquid argon (LAr).  The LAr served as both a coolant and an 
instrumented active shield. It allowed effective detection of the argon scintillation light produced by background events that deposit energy in
the argon surrounding the germanium detectors.  The LAr cryostat was immersed in a 590 m$^3$ water tank equipped with photomultipliers which, 
together with scintillator plates on top of the setup, served as a further shield against external radiation and as a muon veto system. 
In the event of a trigger in one of the HPGe detector channels,
all HPGe readout channels were recorded for off-line analysis. 
$0\nu\beta\beta$ decay candidate events were required to have a point-like energy deposition in a single HPGe detector, and no signal in the 
liquid argon or the muon system.
Based on these selection criteria, a quasi-background free search was performed with a total exposure in detector mass accumulated over \pt\ 
of 103.7~\kgy\ and, after combination with \phaseone , a limit of on the half-life of \onbb\ decay in $^{76}$Ge was set to 
$T_{1/2} > 1.8 \times 10^{26}$ at 90\% C.L. \cite{GERDA:2020xhi}.

Previous simulations \cite{gerda:pandola:2007} identified the delayed decays of $^{77}$Ge and its isomeric state $^{77\mathrm{m}}$Ge, both produced
 by neutron capture on $^{76}$Ge, as the dominant cosmogenic backgrounds in \gerda . 
The Q$_\beta$ value of $^{77}$Ge ($\SI{2703}{keV}$) and of the isomeric state are both above the Q$_{\beta\beta}$ value of 
$^{76}$Ge (\SI{2039}{keV}). 
Thus, their $\beta$ decays can deposit energy in the region of interest at Q$_{\beta\beta}$ and mimic a signal-like event. 
Conversely, $^{77}$As, the decay product of $^{77}$Ge, does not contribute to the background, since with $Q_\beta = \SI{684}{keV}$ it cannot 
contribute in the region of interest.

Recently, a $^{77\mathrm{(m)}}$Ge\footnote{The notation $^{77\mathrm{(m)}}$Ge represents both \germg\ and $^{77\mathrm{m}}$Ge.} production rate of 
(0.21\,$\pm$\,0.01 \newline {(stat)}$\,\pm$\,0.07{(sys)})\,nuc/(kg$\cdot$yr) was obtained using a full \geant\ simulation 
\cite{gerda:virtual:2018}. 
The systematic uncertainties were $\SI{35}{\%}$, dominated by muon-induced neutron production and propagation.
At this rate, the $^{77\mathrm{(m)}}$Ge background contribution at Q$_{\beta \beta}$ is estimated to be 
$\sim 10^{-5}$ \ctsper \ after applying the standard cuts used in \gerda . With the addition of a delayed coincidence cut 
as defined in the above paper, a background contribution of $(2.7\pm 0.3) \times 10^{-6}$ \ctsper \ can be achieved. 
Using this as a proxy for the planned $^{76}$Ge experiment \Lk\ at \lngs\, this showed that the $^{77\mathrm{(m)}}$Ge background contribution can 
be suppressed low enough to achieve a background at Q$_{\beta \beta}$ of $<10^{-5}$ \ctsper , which is a factor of 50 
reduction with respect to \gerda\ \cite{L1K:pCDR:2021}.
Experimental validation of the predicted $^{77\mathrm{(m)}}$Ge production rate and constraining uncertainties using \gerda\ data are therefore of 
paramount importance to consolidate the background model of \Lk . 

This paper summarizes the present analysis to quantify the in-situ production of $^{77}$Ge in its ground state ($T_{1/2} = 11.21$~h 
\cite{ENSDF:As77:2020}) in \gerda\ by searching for its characteristic decay through the isomeric state 
($9/2^+ $, 475\;keV, ${T_{1/2}} = 114 \, \mu $s) of its progeny $^{77}$As. The analysis uses the full data set of the \gerda\ \pt\ experiment with 
an exposure of 103.7~\kgy . With the production rate estimate from above, this gives an expected number of about 
$(22\pm 7)$ $^{77\mathrm{(m)}}$Ge nuclei in either its ground state or isomeric state.  

\begin{figure}  
\begin{minipage}{\columnwidth}
\centering
\includegraphics[width=\columnwidth]{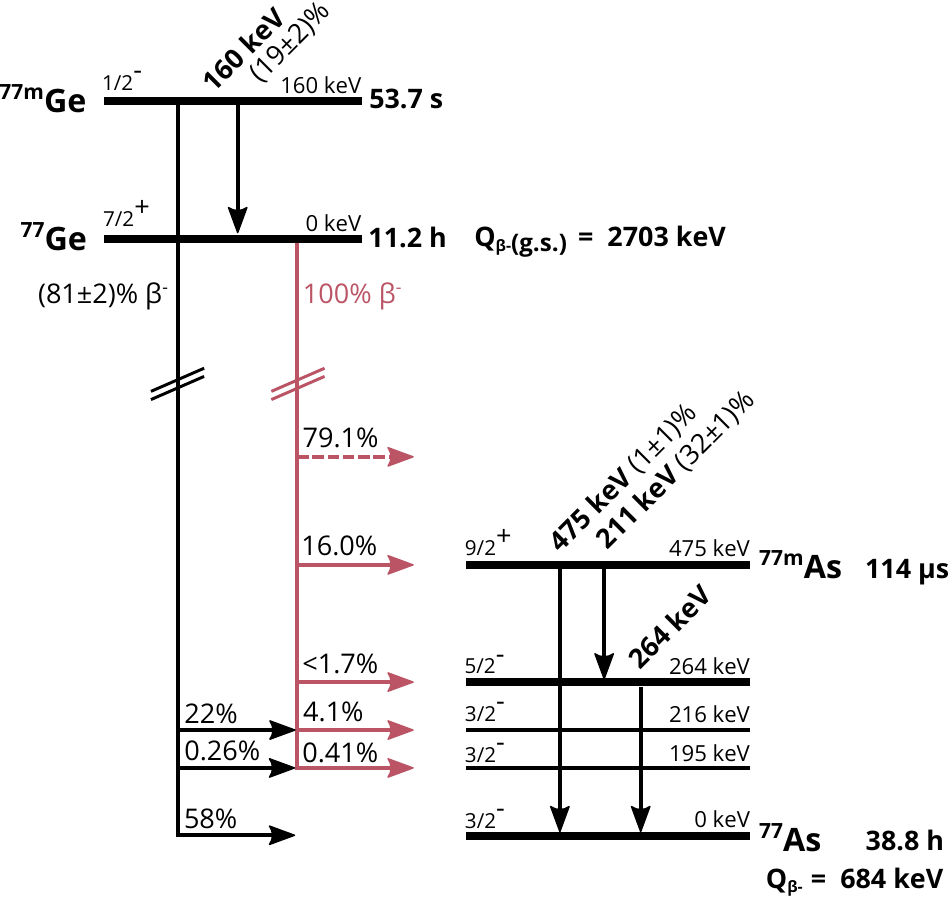}
\end{minipage}
\caption{
Simplified decay scheme of $^{77}$Ge and $^{77\mathrm{m}}$Ge into $^{77}$As. $^{77\mathrm{m}}$Ge only populates states $\leq\SI{216}{keV}$ 
while $^{77}$Ge also populates higher states including the  $9/2^+$ isomeric  state in $^{77}$As with ${T_{1/2}}=\SI{114}{\mu s}$ and an excitation 
energy of $\SI{475}{keV}$.  Approximately $\SI{16}{\%}$ of $^{77}$Ge decays into this isomeric state, while more than $\SI{79}{\%}$ of
$^{77}$Ge decays populate states above (not drawn) and can also populate the isomeric state from above. This plot was generated from the values
of \cite{ENSDF:As77:2020}.
}
\label{fig:my_own_simplified_transition_plot_open_sans}     
\end{figure}

To isolate the $^{77}$Ge decays, we search for a coincidence between the prompt $\beta$ decay of $^{77}$Ge and the delayed de-excitation of the 
isomeric $^{77\mathrm{m}}$As state. We require that the prompt energy deposition and the delayed de-excitation occur in the same detector. 
Figure~\ref{fig:my_own_simplified_transition_plot_open_sans} displays a simplified decay scheme of the $^{77}$Ge-$^{77}$As system. The time 
correlated signature of the beta decay into $^{77\mathrm{m}}$As and its delayed de-excitation leads to so-called pile-up signals in the \gerda\ data
 stream.

Special processing of the \gerda\ data was performed to isolate time-correlated candidates since they were discarded in the standard \gerda\  
$0\nu\beta\beta$ decay a\-na\-ly\-sis. 
Furthermore, new digital signal processing (DSP) routines were developed to extract the physical parameters of the pile-up signal. The DSP routines 
have been validated using generated data and the candidate event selection efficiencies and the uncertainties in the energy reconstruction were 
determined. 

A similar analysis searching for the production of $^{77}$Ge using the delayed de-excitation of the isomeric $^{77\mathrm{m}}$As state was recently 
performed on the \majorana\ demonstrator experiment data \cite{Majorana:2021lgr}.
Considering the larger overburden of the Sanford Underground Research Facility (SURF) at 4.3 km water equivalent (w.e.) compared to LNGS at 3.5 km w.e., 
the production rate of $^{77}$Ge is expected to be lower in comparison.
This analysis found no candidate event for this mechanism, but for other cosmogenic isotopes it found a factor of 2 agreement between simulation and
data.
However, due to the different shielding material, i.e. lead in \majorana\ and LAr in \gerda, it is difficult to use these results to validate 
the \gerda\  simulation, which further motivates the work presented in this paper.


\section{Analysis procedure}
\label{sec:Analysis_procedure}

We distinguish two classes of transitions: \textit{prompt} transitions ($\beta $ and subsequent prompt $\gamma $ de-excitations), shown in 
Figure~\ref{fig:my_own_simplified_transition_plot_open_sans}, start from the ground state of \germg\ and end on the 475~keV 
(${T_{1/2}} = 114 \, \mu $s) isomeric state of $^{77\mathrm{m}}$As. 
\textit{Delayed} transitions, are those electromagnetic de-excitations that start from the 475~keV isomeric state and terminate at the ground state 
of $^{77}$As. 
Since the half-life of $^{77\mathrm{m}}$As is longer than the charge collection time in \gerda's HPGe detectors  
($\lesssim \SI{1.5}{\mu s}$), the delayed transitions occur with sufficiently large time differences with respect to the prompt so that both 
transitions can be separated unambiguously in time.  
We call their combined occurrence in the same detector a \textit{delayed coincidence}.

In Subsection~\ref{sec:Signature_in_GERDA} we identify the expected signature of the delayed coincidence in the \gerda\ experiment. Then, in 
Subsection~\ref{sec:Pile-up_event_reconstruction} we present the DSP to reconstruct pile-up signals in the \gerda\ data and how we estimate their 
energies. Finally, in Subsection~\ref{sec:Candidate_selection} we present the selection criteria to identify candidate delayed coincidences in 
the \gerda\ data and calculate the total selection efficiency. 

\subsection{Signature in GERDA}
\label{sec:Signature_in_GERDA}

In-situ cosmogenic interactions can produce both \germg\ and $^{77\mathrm{m}}$Ge.
The isomeric state $^{77\mathrm{m}}$Ge 
undergoes an internal transition to the ground state \germg\ with $(19\pm2)\%$. When $^{77\mathrm{m}}$Ge decays directly to $^{77}$As, it has
 a $\SI{99}{\%}$ probability of populating one of the four states that are energetically below the isomeric state $^{77\mathrm{m}}$As. 
Therefore, $^{77\mathrm{m}}$Ge decays cannot be tagged through the $114 \, \mu $s delayed coincidence.
Conversely, {$(33\pm1)\%$} of \germg\ decays populate the isomeric state $^{77\mathrm{m}}$As (see 
Figure~\ref{fig:my_own_simplified_transition_plot_open_sans}). 
About half of the prompt decays ($(16\pm1)\%$) that populate the isomeric state are direct transitions, while the other half come from consecutive 
gammas de-excitations from higher levels of $^{77}$As. 
{The end-point energy of these decays into the isomeric state is $\SI{2228}{keV}$ ($\SI{2703}{keV}-\SI{475}{keV}$).} 
The rest of the beta decay branches (\SI{67}{\%}) populate other states that miss the isomeric state in the consecutive gamma decay. 

While the beta particle will deposit its energy in the germanium detector where the $^{77\mathrm{(m)}}$Ge is produced, the gammas can also escape 
and deposit their energy in the liquid argon (LAr) or in other detectors. 
To simulate the prompt transitions of the \germg\ decay into $^{77\mathrm{m}}$As in the \gerda\ \pt\ experiment, we used the MaGe simulation 
framework \cite{gerda:Boswell:2011}. We define the multiplicity as the total number of detectors with coincident energy deposition above 
$\SI{40}{keV}$. At least one of these detectors must have an energy deposition above $\SI{200}{keV}$. 
This additional condition avoids systematic uncertainties in modeling the online trigger threshold, which had different values during different data
 acquisition periods but was always set well below the above value. 
We find that {$>75\%$} of prompt transitions occur with a multiplicity of one, i.e., all energy is deposited in a single detector.

\begin{figure}
\begin{minipage}{\columnwidth}
  \includegraphics[width=\columnwidth]{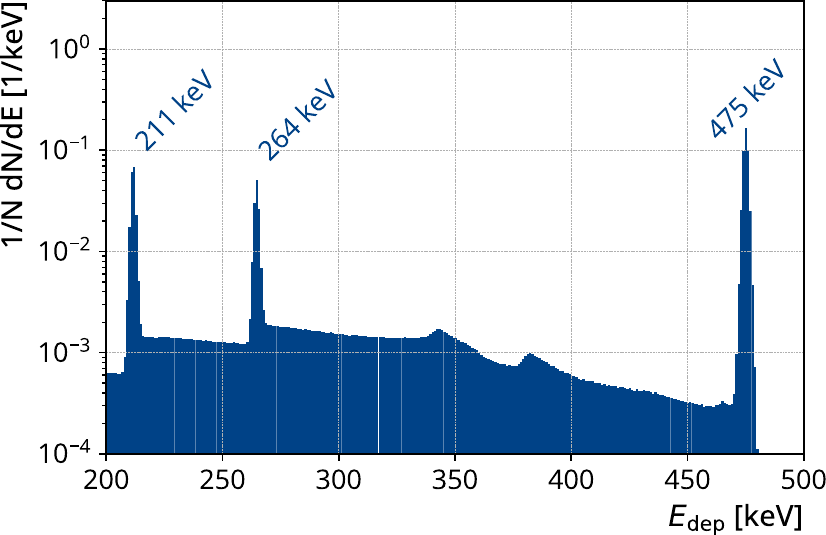}
\end{minipage}
\caption{
Simulated energy distribution of the delayed gamma emission from the isomeric state in $^{77}$As deposited in a HPGe. 
The peaks correspond to the full energy deposition of both gammas in the detector at $\SI{475}{keV}$ or the full energy deposition of only one of 
the gammas in the detector at $\SI{211}{keV}$ or $\SI{264}{keV}$. The energy resolutions of each individual detector derived from the 
standard \gerda\ analysis were implemented.
}
\label{fig:ed_plot}     
\end{figure}

The delayed transitions were also simulated with MaGe. 
We found that $\SI{65}{\%}$ of the energy is deposited only in the same detector as the prompt transition.
Figure~\ref{fig:ed_plot} shows the energy deposited in an HPGe of a delayed transition with multiplicity one. The spectrum shows full energy peaks 
at $\SI{211}{keV}$, $\SI{264}{keV}$, and at $\SI{475}{keV}$ due to the summation of the two gammas or the single gamma transition 
(see Figure~\ref{fig:my_own_simplified_transition_plot_open_sans}). 

\begin{figure}
\begin{minipage}{\columnwidth}
  \includegraphics[width=\columnwidth]{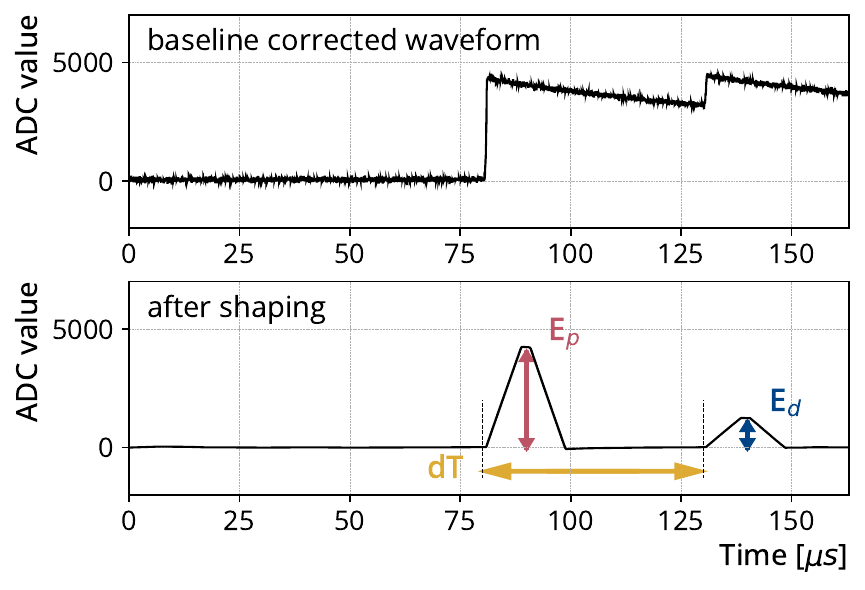}
\end{minipage}
\caption{An example of pile-up signal reconstruction. (Top) Example of a pile-up event waveform in the \gerda\ data stream. (Bottom) The waveform of 
the example pile-up event after applying a trapezoidal filter. The time difference between the signals is estimated by taking the difference between 
the triggers. Finally, the signal heights are extracted with a fixed time pick off.}
\label{fig:dsp_example}     
\end{figure}

{ \gerda\ used charge-sensitive preamplifiers with RC-feedback that shape the physical signal from a HPGe detector into a pulse consisting of a 
rapid change in output voltage followed by a slow exponential decay ($\tau \sim 150\;\mu$s) back to the baseline.} 
The time difference between the prompt and delayed transitions is of the same order as the decay constant of the charge sensitive amplifier. 
Therefore, in the previously mentioned most likely case where the delayed transition only deposits energy in the same detector, the signal produced 
by the delayed transition will appear as a pulse that lies on top of the pulse produced by the prompt transition also known as pile-up 
(e.g., see Figure~\ref{fig:dsp_example}, top).

An energy deposition in any of the 40(41) HPGe detectors generated a synchronous readout of the waveforms of all HPGe detectors. These coincident 
waveforms are referred to as one event. The length of the waveforms is $\SI{164}{\mu s}$, with the initial trigger centered at about 
$\SI{80}{\mu s}$. Therefore, a pile-up signal can occur either in the same waveform as the prompt signal (as see the example 
Figure~\ref{fig:dsp_example}) or in a waveform of a subsequent event. In the \gerda\ experiment, the data acquisition (DAQ) system is configured to 
record a new event if a trigger occurs more than \SI{50}{\mus} after the previous one, with the second waveform centered at 
\SI{80}{\mus}. If the interval between two triggers is shorter than the full waveform duration (\SI{164}{\mus}), the corresponding waveforms will 
overlap, meaning they share a common set of samples. Since the standard \gerda\ analysis rejects overlapping waveforms, we have implemented a new 
DSP tool to accurately reconstruct pile-up signals also for overlapping waveforms.

\subsection{Pile-up signal reconstruction}
\label{sec:Pile-up_event_reconstruction}

The new DSP is based on the validated {Majorana/ GERDA Data Objects Library (MGDO)}, which are also used in the standard \gerda\ 
DSP \cite{gerda:mgdo:2012}. {The DSP transforms the original waveform with trapezoidal filters consisting of a pole-zero correction, a moving window 
differentiation, and a moving window average filter.} A selection of filter lengths was chosen that are optimized for energy resolution and 
detection sensitivity. 
The filter length for a given pile-up signal is selected in steps according to the minimum time difference between the triggers. 
For example, for time differences $\SI{>18}{\mu s}$, the moving window differentiation is $\SI{10}{\mu s}$ wide and the moving window average 
is $\SI{8}{\mu s}$, resulting in a total filter length of $\SI{18}{\mu s}$ and a flat top of $\SI{2}{\mu s}$.
The time difference between signals in the same waveform is calculated as the difference between trigger positions. For pile-up signals in different 
events we extract the time difference using their timestamp in ns. The height of the signals are extracted with a fixed time pick-off at 
the $\SI{75}{\%}$ point on the flat top. Figure~\ref{fig:dsp_example} bottom shows an example of the DSP, extracting the peak heights and time 
difference between the signals.
The DSP also extracts additional quality parameters from the waveforms such as a negative trigger to reject electromagnetic noise and non-physical 
pulses.
We ran the new DSP on waveforms containing two pulses, extracting the heights of both the prompt and the delayed event, and on waveforms containing 
only the delayed pulse. For all other situations, we found that the original DSP gave similarly good results, and we used its energy estimation.

To calibrate the signal heights extracted from the waveform with the new DSP, we use the waveforms of signals with already estimated energies from 
the standard \gerda\ analysis. We apply the same trapezoidal filter to 50 consecutive single signal waveforms after the delayed coincidence 
candidate in the same detector. We plot their signal height against their previously estimated energy, associate them with an uncertainty 
corresponding to the resolution of the detector at that energy  and extract the calibration by fitting the plot with a linear function. 
This gives us the calibrated reconstructed prompt ($E_\mathrm{p}$) and delayed ($E_\mathrm{d}$) energies.

\begin{figure*}[t]
  \includegraphics[width=\textwidth]{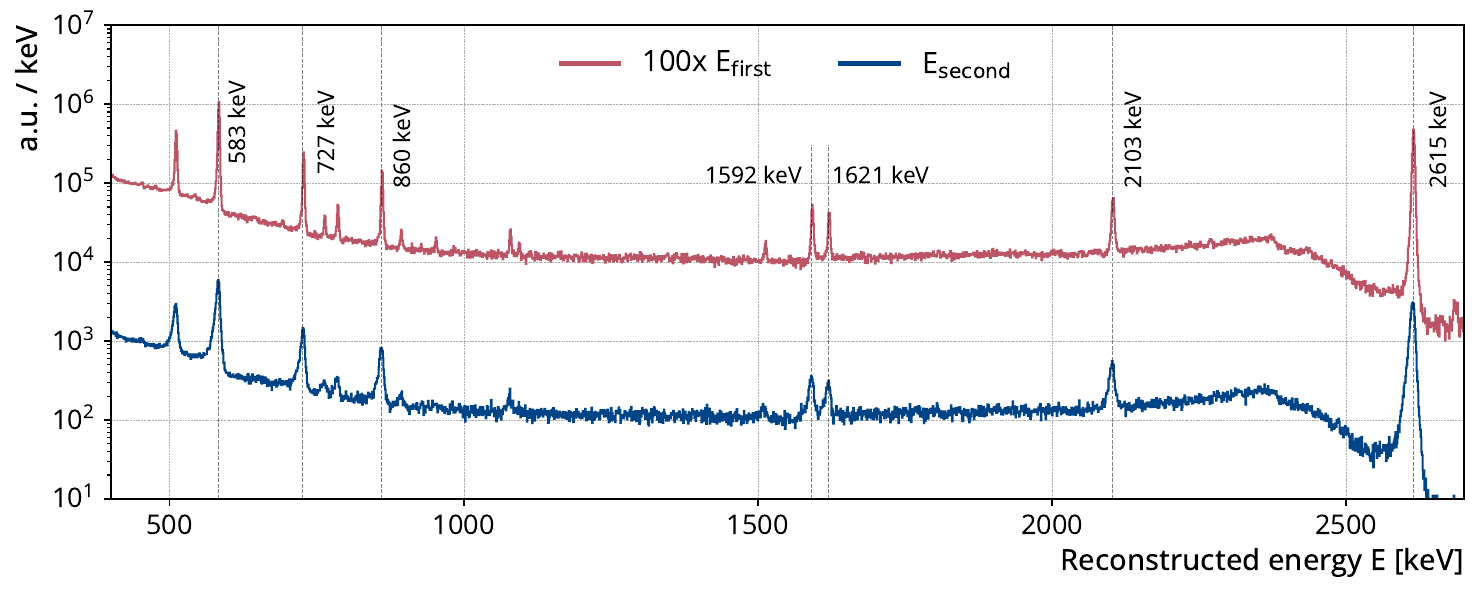}
\caption{Reconstructed energy spectra for the first and second pulse in a pile-up signal in the calibration data where two signals are contained 
in the waveform of one event. The red histogram corresponds to the first pulses energies, while the blue histogram corresponds to the second pulses 
energies. The former distribution was scaled by a factor of 100 to illustrate the differences between the spectra. Both spectra reconstruct the 
expected $^{208}$Tl gamma lines well. The spectra of the second pulses consistently shows larger resolution and a slight tail to lower energies.}
\label{fig:c1_calibration_energies}      
\end{figure*}

Since this is a non-standard approach to energy calibration, we performed several cross-checks. 
In \gerda , to obtain a calibration, dedicated calibration runs were performed before and after each physics run \cite{gerda:calib:2021}. 
During those, three $^{228}$Th sources were automatically lowered into the LAr cryostat in close proximity to the HPGe detectors. 
The calibration data are populated with a large contribution of pile-ups from the gammas emitted by the progeny $^{208}$Tl. 
Since these signals are not correlated, they are called first and second instead of prompt and delayed pulse. 
Figure~\ref{fig:c1_calibration_energies} shows the energy distribution of the first and second pulse energies of pile-up signals in the calibration 
data, where the two signals are contained in the waveform of one event. We can reconstruct the expected gamma lines for both pulses in a pile-up 
signal. The energy distribution of the first pulse shows energy resolutions of e.g. a Full Width Half Maximum (FWHM) of \SI{4.2}{keV} 
at \SI{2.6}{MeV}. Compared to the exposure weighted resolution in the standard \gerda\ analysis of about \SI{3.7}{keV} at \SI{2.6}{MeV} this value 
is slightly higher. In contrast to the standard GERDA analysis, the DSP parameters in this study were optimized for pile-up recognition rather 
than for achieving optimal energy resolution. The energy distribution of the second pulse exhibits a significant degradation in resolution. 
For instance, a FWHM of \SI{8.4}{keV} at \SI{2.6}{MeV} is roughly twice that of the first pulse. This degradation arises because the second pulse 
occurs on the falling tail of the first pulse, resulting in suboptimal baseline reconstruction. In addition, the gamma lines exhibit low-energy 
tailing and a slight energy shift of approximately \SI{-1}{keV} at \SI{2.6}{MeV}, here referred to as the energy bias 
(see Fig. \ref{fig:c1_calibration_energies}).

To further investigate the energy resolution and bias of individual candidates delayed coincidences, we generated data with similar signatures. 
We chose single pulse waveforms from the calibration data with similar prompt energy to the candidate prompt pulse. 
Then we selected another single pulse waveform with energies near the $^{208}$Tl gamma lines as the delayed pulse and added them with a time 
difference similar to the candidate delayed coincidences. Subsequently, we reconstructed the energy of the generated pile-up signal with the DSP 
and extracted the energy resolution and bias from the reconstructed gamma lines by fitting the peaks with a Gaussian peak, a linear background, 
a step function, and a tail function similar to \cite{gerda:calib:2021}. We also modeled the energy resolution in a similar way by taking the 
square root of a linear function and fitting it to the peaks full-width-at-tenth-maximum (FWTM). To obtain the energy resolution of the candidate 
pile-up signal, we interpolated this model at the reconstructed delayed energy. We defined the bias of the candidate pile-up signal energy as the 
largest difference between the reconstructed peak mean energy and the expected energy of all gamma lines. We found that the bias is the largest for 
delayed coincidences with short time differences but usually well below $\SI{1}{keV}$. With this we estimated a FWTM and bias range for each 
candidate delayed coincidence.

\subsection{Candidate selection}
\label{sec:Candidate_selection}

Based on the simulation results, we define the selection criteria for the prompt and delayed energy and the multiplicity as well as the time 
difference of the candidates.

We define a selection condition for prompt transition candidates by requiring that they have (i) a multiplicity of one and (ii) an energy 
$E_\mathrm{p}$ between $\SI{200}{keV}$ and {$\SI{2228}{keV}$ equal to the maximum total deposited energy. The simulation shows that 
$\SI{74.6}{\%}$ of prompt transitions satisfy these conditions.

We further define a delayed transition selection condition that candidate delayed transitions must (i) have multiplicity one, (ii) occur in the same 
detector as the prompt transition, and (iii) have one of the three gamma energies ($\SI{211}{keV}$, $\SI{264}{keV}$ or $\SI{475}{keV}$) lie within 
the respective reconstructed energy acceptance region $E_\mathrm{d}$ defined by the FWTM range plus bias. We chose to use the FWTM (containing 
$\SI{96}{\%}$ of the peak area) rather than the FWHM to account for possible tails as gamma lines in the energy spectrum of the delayed candidate 
signals may deviate from the expected Gaussian distribution (see Figure~\ref{fig:c1_calibration_energies}). To account for a potential bias in the 
reconstructed delayed energies, we extend the acceptance region of an individual delayed coincidence candidate a\-sym\-me\-tri\-cal\-ly by the 
minimum and maximum bias described in Subsection~\ref{sec:Pile-up_event_reconstruction}. 
We find that $\SI{46.2}{\%}$ of the delayed transitions satisfy the above conditions. The energy and multiplicity selection efficiency of 
$\SI{34.5}{\%}$ in Table~\ref{tab:efficiencies} is the combined efficiency to select both the prompt and the delayed transition with these conditions. 

Finally, we require that the time difference $dT$ bet\-ween a prompt and a delayed transition candidate be no greater than five times the lifetime 
of the isomeric state ($\SI{822}{\mu s}$). The corresponding time difference selection efficiency is also given in Table~\ref{tab:efficiencies}.

We now define the criteria for selecting the candidates based on the quality parameters extracted from the standard \gerda\ analysis as well as the 
pile-up reconstruction using our DSP. Due to the difference in signature between delayed coincidences with signals in the same or different 
waveforms, we define three different regions depending on the time difference between the signals.

1. For time differences less than $\SI{70}{\mu s}$, the two signals will be contained in one waveform. This is because the first signal is 
approximately in the center at $\SI{80}{\mu s}$ of the long $\SI{164}{\mu s}$ waveform. 
We require that the quality parameters that are extracted from the new DSP satisfy a set of conditions. 
A candidate for a delayed coincidence with corresponding time differences must have (i) exactly two triggers and (ii) pass through the additional 
quality parameter cuts.

2. For time differences between $\SI{70}{\mu s}$ and $\SI{164}{\mu s}$, the two signals are contained in separate events, but are still so close 
together that their waveforms overlap. Since the DAQ records a new event when a trigger occurs at least $\SI{50}{\mu s}$ after the previous one, 
we are sensitive to all signals in that time range.
In the standard \gerda\ analysis, the second event was discarded before the DSP step. This is due to a short busy period of the DAQ of the order 
of 100\,ns during the saving process of the first event, which interrupted the recording of the waveform of the following event. 
We reconstructed such waveforms by interpolating in these periods. Since these events were not processed in the standard \gerda\ analysis, we 
applied DSP to all its waveforms. Using the extracted quality parameters, we require that a candidate delayed signal waveform with corresponding 
time difference (i) contains only one trigger and (ii) pass through the additional quality parameter cuts. 

3. For time differences above $\SI{164}{\mu s}$, the waveforms of the two events no longer overlap, and both events have been processed in the 
standard \gerda\ analysis. We use the standard \gerda\ quality conditions to differentiate signal candidates from non-physical signals. We require 
that the prompt waveform satisfy all standard \gerda\ quality conditions, while the delayed waveform satisfies a subset of these conditions 
consistent with a signal on an exponential tail. 
For these events, we did not apply the new DSP to either the prompt or the delayed waveform because the standard \gerda\ analysis already provided 
reliable values of the prompt and delayed energies.

To estimate the resulting pile-up signal selection efficiency, we generated pile-up signals with energies in the corresponding energy ranges and 
time differences sampled from an exponential distribution according to the lifetime of $^{77\mathrm{m}}$As. We define the pile-up signal selection 
efficiency as the number of signals that pass the entire pile-up signal selection procedure over the total amount generated. The final pile-up 
signal selection efficiency is given in Table~\ref{tab:efficiencies}. We tested whether the pile-up signal selection efficiency depends on certain 
parameters. Figure~\ref{fig:event_acceptance} shows the pile-up signal selection efficiency plotted over the time difference range up to 
$\SI{20}{\mu s}$. It shows a jump from zero to almost one at $\SI{4.5}{\mu s}$, which shows that our analysis is sensitive to pile-ups with time 
differences above this threshold.  The efficiency of pile-up signal selection remains constant above $\SI{20}{\mu s}$ up to the upper limit of time 
difference selection at $5\,\tau_\mathrm{^{77m}As} = \SI{822}{\mu s}$. Therefore, the only dead time in our analysis is due to the delayed signals 
that occur at time differences smaller than $\SI{4.5}{\mu s}$. We also tested for a prompt or delayed energy dependence and found it to be constant 
over the ranges of interest.
    
The total selection efficiency 
$\epsilon_{\mathrm{total}} = \epsilon_{\mathrm{em}} \times \epsilon_{\mathrm{dT}} \times \epsilon_{\mathrm{pile-up}}$ amounts to $32.4\%$.

\begin{figure}
  \includegraphics[width=\columnwidth]{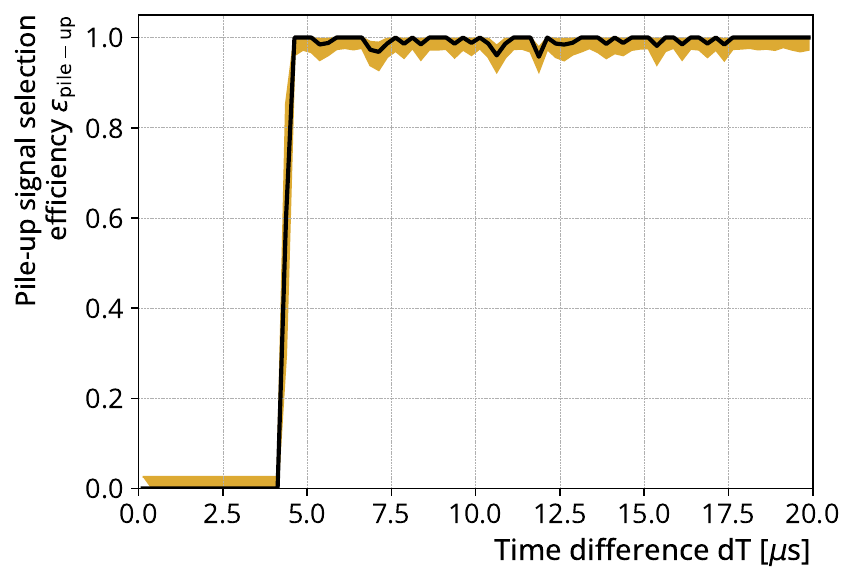}
\caption{Pile-up signal selection efficiency plotted over the time difference between generated pile-up signals. Our new DSP is sensitive for 
delayed coincidences starting at time differences $\SI{>4.5}{\mu s}$. Individual events in this range were rejected due to quality cuts. 
Black: central value. Yellow: 68\% uncertainty band.}
\label{fig:event_acceptance}     
\end{figure}

\begin{table}

\caption{Delayed coincidence selection efficiencies. The uncertainty of these values is in the order of $\SI{<0.5}{\%}$.\label{tab:efficiencies}}
\begin{tabular}{lr}

\hline\noalign{\smallskip}
Energy and multiplicity selection efficiency $\epsilon_{\mathrm{em}}$ & \hspace{0.125cm}$34.5\%$\\
\hspace{1cm} {\scriptsize prompt contribution $\epsilon_{\mathrm{em}}^\mathrm{p}$} & {\scriptsize$74.6\%$}\\
\hspace{1cm} {\scriptsize delayed contribution $\epsilon_{\mathrm{em}}^\mathrm{d}$} & {\scriptsize $46.2\%$}\\
Time difference selection efficiency $\epsilon_{\mathrm{dT}}$ & \hspace{0.125cm}$99.3\%$\\
Pile-up signal selection efficiency $\epsilon_{\mathrm{pile-up}}$& \hspace{0.125cm}$94.9\%$ \\
\noalign{\smallskip}\hline
Total selection efficiency $\epsilon_{\mathrm{total}}$ & \hspace{0.125cm}$32.4\%$\\
\noalign{\smallskip}\hline

\end{tabular}
\end{table}

\begin{figure*}
  \includegraphics[width=\textwidth]{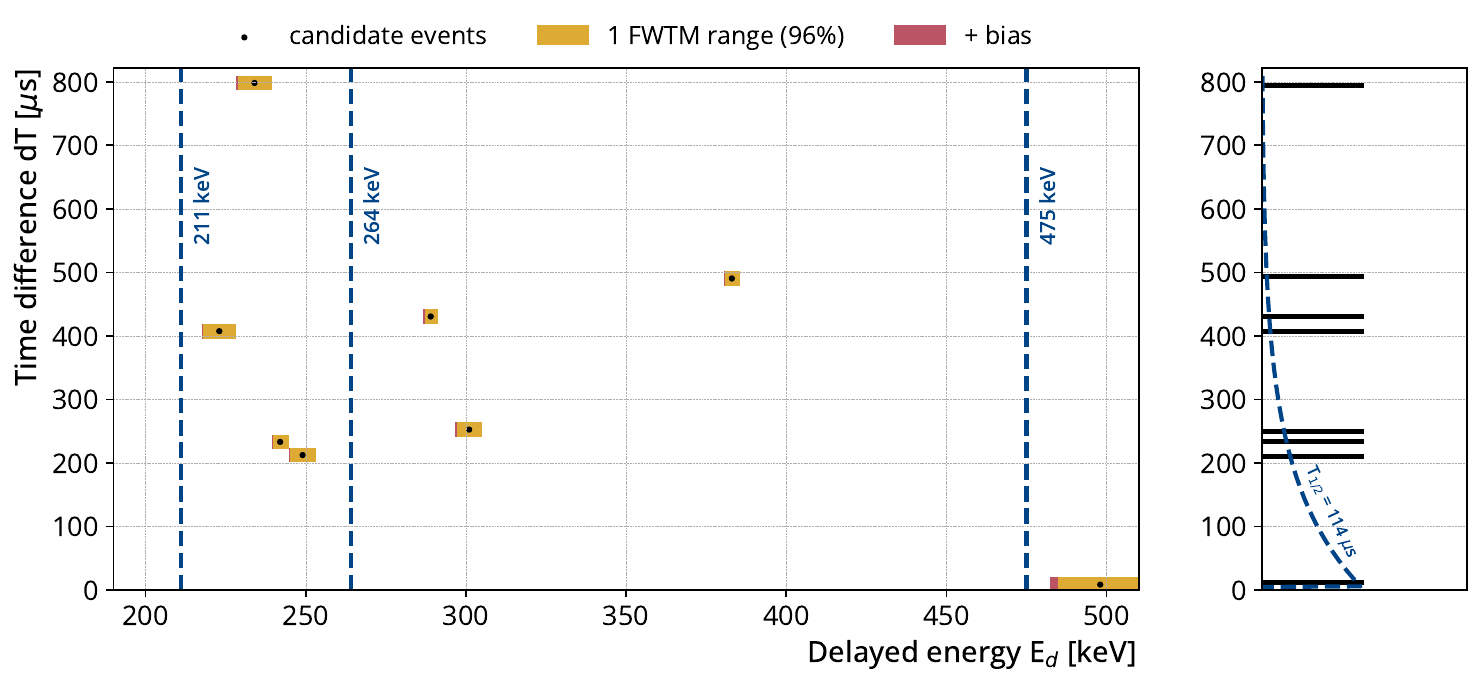}
\caption{The distribution of all 8 candidate delayed coincidences after the multiplicity condition plotted $dT$ over $E_\mathrm{d}$. The energy 
windows around the points correspond to the linear combination of the full-width-at-tenth-maximum (FWTM) window (yellow) plus the asymmetric 
bias window (red). The FWTM window covers about $97\%$ of the gamma peak area. The vertical size of the data is enlarged for better visualization. 
The blue lines correspond to the gamma energies of the internal transitions from $^{77\mathrm{m}}$As . A delayed coincidence candidate is rejected, 
if its energy window misses any of the three gamma lines.
We found no candidate delayed coincidence that satisfies this condition resulting in $N_\mathrm{obs} = \SI{0}{cts}$. The right part shows a 
projection of the candidates onto $dT$. The blue line corresponds to the expected distribution for $^{77\mathrm{m}}$As delayed coincidences.}
\label{fig:scatter_plot}      
\end{figure*}


\section{Result}
\label{sec:result}

We applied the pile-up signal selection to the entire \gerda\ \pt\ data set, which has an exposure of $103.7$~\kgy. Figure~\ref{fig:scatter_plot} 
shows a scatter plot of the delayed coincidence candidates that pass all the selection criteria of the previous section apart from the delayed 
energy selection.
The black dots represent the reconstructed values of delayed energy and time difference, while the colored bars in x-direction represent the delayed 
energy FWTM plus bias acceptance region (linearly summed) estimated individually for each candidate. A candidate is accepted if its reconstructed 
delayed energy acceptance region overlaps with one of the expected gamma energies. We find that there is no candidate delayed coincidence that 
passes this selection ($N_\mathrm{obs} = \SI{0}{cts}$). 

Random coincidences are the only known background source in this analysis. Using the average signal rates in each detector for the 1 FWTM energy 
ranges around the gamma lines, we calculated the number of expected random coincidence signals considering our selection criteria to be 
$N_\mathrm{rc} = \SI{0.04}{cts}$.

For a wider range of delayed energies between $\SI{200}{keV}$ and $\SI{500}{keV}$ the expected number of random coincidence signals is 
$\SI{0.2}{cts}$ compared to eight delayed coincidence candidates observed (see Figure~\ref{fig:scatter_plot}). The time and energy distributions of 
these signals differ significantly from the expected \germg\ delayed coincidence time and energy distributions, and thus can be excluded as a 
majority contributor to these events.
If we ignore the time distribution, for the \germg\ delayed coincidences that deposit energy above $\SI{200}{keV}$, the fraction of energy 
depositions in the continuum region is around $\SI{26}{\%}$, with the rest in the peaks.
Without an event in the peaks, it is therefore feasible that one or two events in the continuum could be \germg\ delayed coincidences, but more 
would be unlikely.
A search for other sources generating delayed coincidences such as delayed neutron capture in muon-induced showers, $^{214}$Bi-$^{214}$Po decays, 
or delayed de-excitation in other isotopes, was performed but was inconclusive.

With $N_\mathrm{obs} = \SI{0}{cts}$ and $N_\mathrm{rc} = \SI{0.04}{cts}$, we can determine an upper limit of $<\SI{2.4}{cts}$ ($90\%$ CL) on the 
number of delayed coincidences by applying Feldman-Cousins method \cite{stat:Feldman:1997qc}. We can convert the number of delayed coincidences to 
the number of \germg\ decays by dividing it with the product of the total selection efficiency and the ratio of delayed coincidences per \germg\ 
decay ({$(33\pm1)\%$}, see Figure~\ref{fig:my_own_simplified_transition_plot_open_sans}). We treat the uncertainty by integrating over the nuisance 
parameter as described in \cite{stat:CousinsHighland:1992,stat:pdg:2022}. We found an upper limit on the number of \germg\ decays of 
{$N_\mathrm{^{77}Ge} = \SI{22.4}{decays}$ ($90\%$ CL)}. 

Considering that the \germg\ has a half-life much shorter than the time during which  \gerda\ \pt\ acquired the 103.7~\kgy\ exposure, any  \germg\ 
must have been produced in-situ. We can therefore calculate an upper limit on the \germg\ production rate of 
{$<\SI{0.216}{nuc/(kg\cdot yr)}$ ($90\%$ CL)}. 

The \germg\ in \gerda\ is mainly produced by cosmogenic activation. Radiogenic neutrons from natural decay chains originating outside the experiment 
are absorbed by the water tank and cannot reach the HPGe detectors. In the LAr cryostat, a cosmic muon can produce particle showers with high 
neutron multiplicity. The vast majority of neutrons are captured by $^{40}$Ar or by the water in the water tank. A minority is captured by 
$^{76}$Ge, producing either the ground state $^{77}$Ge or the isomeric state $^{77\mathrm{m}}$Ge. The production ratio between the two depends on 
the kinetic energy of the neutron when absorbed. Following the arguments from our previous work \cite{gerda:virtual:2018} and corroborated by 
statistical model calculations \cite{peters_paper}, we assign a probability for the direct ground state population of $\epsilon_{d} = (50\pm10)\%$. 
In addition, the probability of the internal transition from the isomeric state into the ground state of $\epsilon_\mathrm{IT} = (19\pm2)\%$ has to 
be taken into account. Therefore, the probability to populate the ground state $\epsilon_\mathrm{g}$ after neutron capture is  
\begin{center}
$\epsilon_\mathrm{g} = (\epsilon_\mathrm{d} + (1 - \epsilon_\mathrm{d}) \cdot \epsilon_\mathrm{IT}) = (59.5\pm8.1)\%$.
\end{center}
Assuming that all delayed coincidences are of cosmogenic origin, we can use $\epsilon_\mathrm{g}$ to calculate the sum of the $^{77}$Ge and 
$^{77\mathrm{m}}$Ge production rate. We treat the nuisance parameters as before. Applying the Feldman-Cousins method, we get a total 
$^{77\mathrm{(m)}}$Ge production rate of 
\begin{center}
$< {0.38}$\,{nuc/(kg$\cdot$yr)} ($90\%$ CL).
\end{center}

This is the strongest experimental constraint on the $^{77(\mathrm{m})}$Ge production rate at LNGS and is about a factor of {two larger than the 
MC prediction of 
$(0.21 \pm 0.01\mathrm{(stat)}$$\-\pm\- 0.07\mathrm{(sys)})$\,{nuc/(kg$\cdot$yr)}.
 For completeness, we refer here to a previous analysis performed on the \gerda\ data, in which an upper limit of 
$<4.1$\,nuc/(kg$\cdot$yr) ($90\%$ CL) was derived in \cite{gerda:LauraVanhoefer:2018}.

With our constraint on the $^{77\mathrm{(m)}}$Ge production rate, we can update the predictions from previous simulations \cite{gerda:virtual:2018}.
To this end, the simulated production rate estimate is treated as a prior with a Gaussian distribution centered at 
{0.21}\,{nuc/(kg$\cdot$yr)}  
and an uncertainty of 
{0.07}\,{nuc/(kg$\cdot$yr)} ($1 \sigma$).
Our constraint on the $^{77\mathrm{(m)}}$Ge production rate is modeled by an exponential function as illustrated in Fig.~\ref{fig:countur}.
The posterior has a central value of {0.18}{nuc/(kg$\cdot$yr)} with a $1\sigma$ credibility interval of [0.106, 0.251] {nuc/(kg$\cdot$yr)}. 
Relatively, the posterior predicts the simulated production rate scales by a factor of {$0.85^{+0.35}_{-0.34}$} compared to the original estimate.

Previous simulations of the background from $^{77\mathrm{(m)}}$Ge decays in \gerda\ predicted a background index (BI) of 
$(2.7\pm 0.3) \times 10^{-6}$ 
\ctsper \  with individual contributions of 
$(1.2\pm 0.5) \times 10^{-6}$ \ctsper \ and 
$(1.5\pm 0.2) \times 10^{-6}$ \ctsper \ from $^{77\mathrm{m}}$Ge and $^{77}$Ge respectively excluding the previously mentioned 
systematic uncertainties of $\SI{35}{\%}$ \cite{gerda:virtual:2018}.
This BI is achieved after active background suppression (i.e. detector anti-coincidence, rejection by liquid argon anti-coincidence and pulse shape 
discrimination (PSD)), and after applying a veto condition after tagged muons.  
Applying the scaling factor for the $^{77\mathrm{(m)}}$Ge production rate, we estimate a contribution to BI of 
$(2.3\pm0.1\mathrm{(stat)}\pm1.0\mathrm{(sys)}) \times 10^{-6}$ \ctsper .
The statistical uncertainty is due to the finite exposure simulated in the previous simulation study.
The systematic uncertainty consists of the approximately symmetric distribution of the production rate scaling factor with standard deviation 
$\SI{35}{\%}$ and the uncertainty of the direct population of the ground state after production of $\SI{10}{\%}$.

Tagging the delayed coincidence decay through the isomeric state in \nuc{As}{77} can not only be used to estimate the production rate, but also to 
tag and reject  \nuc{Ge}{77} decays in the \onbb\ decay search. To estimate how many such $^{77}$Ge decays can be additionally rejected, we simulated 
$^{77}$Ge decays using a MaGe simulation of the GERDA setup to and modeled the active background suppression in the same way as in 
\cite{gerda:virtual:2018}. 
We found that of the $^{77}$Ge decays with an energy around Q$_{\beta \beta}$, $\SI{92}{\%}$ are rejected by the active background suppression. Of 
the surviving events, about $(62\pm1)\%$ produce a delayed coincidence signal with E$_\mathrm{d} > \SI{200}{keV}$ and $dT > \SI{10}{\mu s}$.
The large contribution of such events can be explained by the fact that $\SI{16}{\%}$ of $^{77}$Ge decay directly into the isomeric state, which are 
simple beta decays with a similar topology to \onbb\ decays and therefore predominantly survive the analysis cuts. By rejecting events with the 
delayed coincidence signal mentioned above, we estimate that a $^{77\mathrm{(m)}}$Ge background index contribution of 
$(1.50\pm0.07\mathrm{(stat)}\pm0.67\mathrm{(sys)}) \times 10^{-6}$ \ctsper \ is achievable. 
The contribution of this rejection to the \onbb\ decay survival fraction is negligible due to the low random coincidence rate.

\begin{figure} 
\begin{minipage}{\columnwidth}
\centering
\includegraphics[width=\columnwidth]{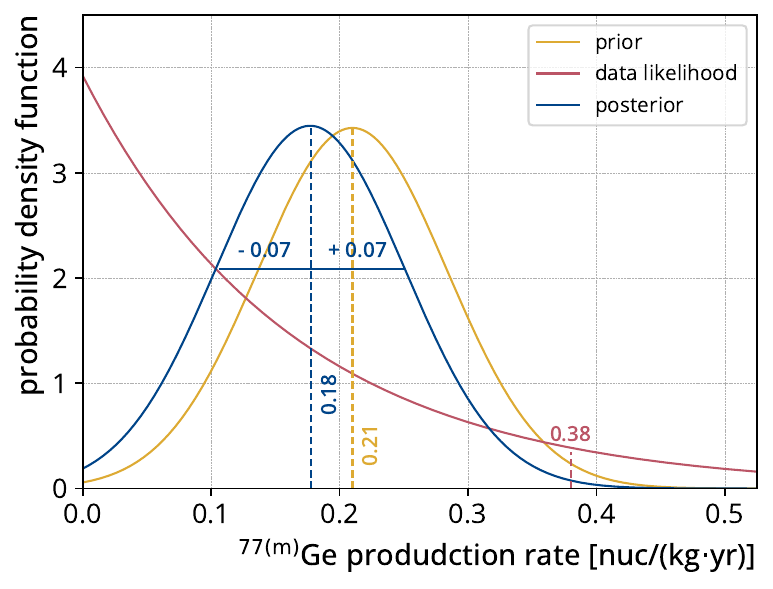}
\end{minipage}
\caption{Bayesian update of the simulated production rate using the likelihood of the \gerda\ data estimated in this analysis.}
\label{fig:countur}     
\end{figure}

\section{Implications and Conclusions}
\label{sec:implications}

LEGEND is the successor experiment to \gerda , which in its first phase, LEGEND-200, reuses the \gerda\ cryostat at \lngs\ and operates up to 200 kg 
of HPGe detectors enriched in the isotope $^{76}$Ge.  In the second phase, \Lk , one tonne of enriched HPGe detectors will be deployed in a new 
experimental infrastructure in Hall C of \lngs . 
After one year of \Ltwo\ data taking with the final detector mass, the sensitivity of the analysis presented here will double, tightening the 
constraint on the $^{77\mathrm{(m)}}$Ge production rate by a factor of two, or potentially revealing a signal.

A previous simulation for \Lk , also based on \geant , gave a $^{77\mathrm{(m)}}$Ge production rate of 
$(0.33 \pm 0.01 \mathrm{(stat)}$$ \pm 0.12 \mathrm{(sys)})$\,{nuc/(kg$\cdot$yr)} \cite{L1K:NeutrinoPoster:2022}.
The approximately 1.6-fold increase in the $^{77\mathrm{(m)}}$Ge production rate is attributed to the larger LAr volume compared to \gerda , which 
results in a higher neutron flux at the HPGe detectors and consequently a greater neutron capture rate. 
As described in \cite{L1K:NeutrinoPoster:2022}, a neutron moderator will be installed around the HPGe strings, dividing the liquid argon (LAr) volume 
into an inner and an outer region. This design aims to moderate neutrons crossing from the outer to the inner region to lower energies where capture 
on $^{40}$Ar is more likely than on $^{76}$Ge. By thus removing the contribution of neutrons originating outside the moderator, the 
$^{77\mathrm{(m)}}$Ge production rate is effectively reduced to a level below that observed in \gerda .
As this decouples the production rate from the size of the cryostat, changes in the ongoing development of the experiment geometry are not expected 
to change this estimate. Furthermore, study \cite{L1K:NeutrinoPoster:2022} estimates that only $\SI{2.3}{\%}$ of all $^{77\mathrm{m}}$Ge decays 
in \Lk\ will survive the veto condition following tagged muons.
This 2.6-fold reduction in the survival fraction, compared to the $\SI{6}{\%}$ observed in \gerda\ \cite{gerda:virtual:2018}, is primarily attributed 
to the tagging of neutrons in liquid argon in hadronic showers, which exhibit high neutron multiplicity. 
Combined with the tagging of the $^{77}$Ge decays transitioning through the isomeric state in \nuc{As}{77}, we estimate that the in-situ cosmogenic 
background contribution for \Lk\ will be below $10^{-6}$ \ctsper .\footnote{
    This estimate is based on the background estimates of \gerda\ for  $^{77\mathrm{m}}$Ge of 
    $(1.2\pm 0.5) \times 10^{-6}$ \ctsper , and for $^{77}$Ge of 
    $(1.5\pm 0.2) \times 10^{-6}$ \ctsper , and the survival fractions for the additional cuts 
    (survival fraction of $^{77}$Ge after $^{77\mathrm{m}}$As decay tagging: $(1-62\%) = 38\%$; survival fraction of $^{77\mathrm{m}}$Ge after muon 
    tagging: $2.3\%$ ($6\%$) in \Lk\ (\gerda ) after scaling by the Bayesian update of the simulation estimate ($0.85$). \newline
    $8.8 \times 10^{-7}$ \ctsper = \newline   
    \phantom\quad\quad\quad  $0.85 \times [ 1.5 \times 10^{-6}$ \ctsper $\times 38\% $\,+\newline
    \phantom\quad\quad\quad  $1.2 \times 10^{-6} $ \ctsper $ \times (2.3\% / 6\%) ].\newline $
    Since the background contribution of $^{77\mathrm{m}}$Ge in \gerda\ already accounted for the survival fraction, the corresponding \Lk\ estimate 
    was scaled according to the better survival fraction. 
    In addition, the background index contribution is expected to be lower in \Lk\ compared to \gerda , as the neutron moderator will reduce the 
    production rate of cosmogenic isotopes. These estimates can also be refined by use of the results from \cite{peters_paper}.
}
The goal of \Lk\ is to achieve a background index of 
$<10^{-5}$~cts/(keV$\cdot$kg$\cdot$yr) for a quasi-background-free search for $0\nu\beta\beta$ decays \cite{L1K:pCDR:2021}. The contribution of 
in-situ cosmogenic background thus accounts for $\SI{\leq 10}{\%}$ of the total background budget. These findings indicate that the rock overburden 
at LNGS, combined with the suppression strategies discussed in \cite{gerda:virtual:2018,L1K:NeutrinoPoster:2022}, and the delayed coincidence method 
to identify $^{77}$Ge decay via the isomeric state of $^{77}$As presented in this paper, are highly effective in reducing the cosmogenic background 
in \Lk\ to a sub-dominant level.

\begin{acknowledgement}
  The \gerda\ experiment is supported financially by the German Federal Ministry
  for Education and Research (BMBF), the German Research Foundation (DFG), the
  Italian Istituto Nazionale di Fisica Nucleare (INFN), the Max Planck Society
  (MPG), the Polish National Science Centre (NCN, Grant number
  UMO-2020/37/B/ST2/03905), the Polish Ministry of Science and Higher Education
  (MNiSW, Grant number DIR/WK/2018/08), the Russian Foundation for Basic
  Research, and the Swiss National Science Foundation (SNF). This project has
  received funding/support from the European Union’s Horizon 2020 research and
  innovation programme under the Marie Sklodowska-Curie Grant agreements no
  690575 and no 674896. This work was supported by the Science and Technology
  Facilities Council, part of the UK Research and Innovation (Grant no.
  ST/T004169/1). The institutions acknowledge also internal financial support.
  The \gerda\ collaboration thanks the directors and the staff of the LNGS for
  their continuous strong support of the \gerda\ experiment.
\end{acknowledgement} 



\begin{thebibliography}{}
%

\bibitem{gerda:phase2:2018}
M. Agostini \etal\ (\gerda\ Collab.), Upgrade for Phase II of the \gerda\ experiment , Eur. Phys. J. C 78, 388 (2018) \url{https://doi.org/10.1140/epjc/s10052-018-5812-2}


\bibitem{GERDA:2020xhi}
{
M. Agostini \etal\ (\gerda\ Collab.), Final Results of \gerda\ on the Search for Neutrinoless Double-$\beta$ Decay, 
Phys. Rev. Lett. 125, 252502 (2020) \url{https://doi.org/10.1103/PhysRevLett.125.252502}
}

\bibitem{gerda:pandola:2007}
L. Pandola \etal , Monte Carlo evaluation of the muon-induced background in the \gerda\ double beta decay experiment, 
Nucl. Instrum. Methods A 570, 149–158 (2007) \url{https://doi.org/10.1016/j.nima.2006.10.103}

\bibitem{gerda:virtual:2018}
C. Wiesinger, L. Pandola and S. Schönert, Virtual depth by active background suppression: Revisiting the cosmic muon induced background of \gerda\ Phase II, 
Eur. Phys. J. C 78, 597 (2018) \url{https://doi.org/10.1140/epjc/s10052-018-6079-3}

\bibitem{L1K:pCDR:2021}
{
N. Abgrall \etal\ (\LEG\ Collab.), The Large Enriched Germanium Experiment for Neutrinoless $\beta\beta$ Decay: LEGEND-1000 Preconceptual Design Report, arXiv (2021)  	
\url{https://doi.org/10.48550/arXiv.2107.11462}
}

\bibitem{Majorana:2021lgr}
I. J. Arnquist \etal , (\majorana\ Collab.), Signatures of muonic activation in the \majorana\ Demonstrator, 
Phys. Rev. C 105, 014617 (2022) \url{https://doi.org/10.1103/PhysRevC.105.014617}

\bibitem{ENSDF:As77:2020}
From ENSDF database as of February 21st, 2024.
Version available at \url{http://www.nndc.bnl.gov/ensarchivals/}
Full Evaluation Balraj Singh ENSDF 30-Sep-2020

\bibitem{gerda:Boswell:2011}
M. Boswell \etal , MaGe – a Geant4-based Monte Carlo application framework for low-background germanium experiments. 
IEEE Trans. Nucl. Sci. 58, 1212 (2011)  \url{https://doi.org/10.1109/TNS.2011.2144619}

\bibitem{gerda:AndreaLazzaro_PhD:2019}
A. Lazzaro, Signal processing and event classification for a background free neutrinoless double beta decay search with the \gerda\ experiment, 
PhD Thesis, Technische Universität München (2019) \url{https://mediatum.ub.tum.de/1507626}

\bibitem{gerda:mgdo:2012}
M. Agostini \etal , The MGDO software library for data analysis in Ge neutrinoless double-beta decay experiments, 
J. Phys. Conf. Ser. 375 042027 (2012) \url{https://doi.org/10.1088/1742-6596/375/1/042027}

\bibitem{gerda:calib:2021}
M. Agostini \etal\ (\gerda\ Collab.), Calibration of the \gerda\ experiment, 
Eur. J. Phys. C 81 (2021) 682 \url{https://doi.org/10.1140/epjc/s10052-021-09403-2}

\bibitem{stat:Feldman:1997qc}
G. J. Feldman and R. D. Cousins, A Unified approach to the classical statistical analysis of small signals, 
Phys. Rev. D 57 (1998) 3873-3889 \url{https://doi.org/10.1103/PhysRevD.57.3873}

\bibitem{stat:CousinsHighland:1992}
R. D. Cousins and V. Highland, Incorporating systematic uncertainties into an upper limit, 
Nucl. Instrum. Meth. A 320 (1992) 331-33 \url{https://doi.org/10.1016/0168-9002(92)90794-5}

\bibitem{stat:pdg:2022}
R. L. Workman \etal\ (Particle Data Group), Prog. Theor. Exp. Phys. 2022, 083C01 (2022)


\bibitem{peters_paper}
P. Grabmayr, Cross sections and gamma cascades in $^{77}$Ge needed for background reduction in $0\nu\beta\beta$ experiments on $^{76}$Ge, 
Eur. Phys. J. A (2024) 60, 115, \url{https://doi.org/10.1140/epja/s10050-024-01336-0}


\bibitem{gerda:LauraVanhoefer:2018}
L. Vanhoefer, Limitations of Rare Event HPGe Experiments due to Muon-Induced Neutron Background, 
PhD Thesis, Technische Universität München (2018) \url{https://mediatum.ub.tum.de/1446626}


\bibitem{L1K:NeutrinoPoster:2022}
M. Neuberger, M. Morella \etal , Strategies for cosmogenic $^{77(\mathrm{m})}$Ge reduction for LEGEND-1000 Experiment, 
Poster, Neutrino 2022 \url{https://doi.org/10.5281/zenodo.6804443}





\end{thebibliography}
\end{document}